# *Premier pas vers un apprentissage supervisé quantique appliqué aux potentiels évoqués cognitifs*

~

# *First steps towards quantum machine learning applied to the classification of event-related potentials*

~

Grégoire Cattan[1], Alexandre Quemy[2], Anton Andreev[3]

[1]IBM Software, Data and AI, Kraków, Poland

[2]Faculty of Computer Sciences, Poznan University of Technology, Poznan, Poland

[3]GIPSA-lab, CNRS, University Grenoble-Alpes, Grenoble INP, Grenoble, France

*Mots-clés : Électroencéphalographie (EEG), Interface Cerveau-Ordinateur (ICO), Interface Cerveau-Machine (ICM), Potentiel Evoqué Cognitifs (PEC), P300, Classification, Apprentissage supervisé, Informatique quantique, Machine a vecteurs de support (SVM)*

**Abstract**— Low information transfer rate is a major bottleneck for brain-computer interfaces based on non-invasive electroencephalography (EEG) for clinical applications. This led to the development of more robust and accurate classifiers. In this study, we investigate the performance of quantum-enhanced support vector classifier (QSVC). Training (predicting) balanced accuracy of QSVC was 83.17 (50.25) %. This result shows that the classifier was able to learn from EEG data, but that more research is required to obtain higher predicting accuracy. This could be achieved by a better configuration of the classifier, such as increasing the number of shots.

**Résumé**— Le faible taux de transfert d'information des interfaces cerveau-machines basées sur l'électroencéphalographie non invasive (EEG) est un obstacle majeur pour les applications cliniques. Cela a conduit au développement de classifieurs plus robustes et plus précis. Dans cette étude, nous étudions les performances d'un classifieur a vecteurs de support quantiquement amélioré (QSVC). La précision équilibrée obtenue lors de la phase d'entrainement (prédiction) avec QSVC était de 83,17 (50,25) %. Ce résultat montre que le classifieur est en mesure d'apprendre à partir des données EEG, mais que des recherches supplémentaires sont nécessaires pour obtenir une plus grande précision de prédiction. Cela pourrait être réalisé par une meilleure configuration du classifieur, comme l'augmentation du nombre d'essais.

# Introduction

Les potentiels évoqués cognitifs (PEC) sont des potentiels de faibles amplitudes produits par le cerveau suite à une stimulation. Les interfaces cerveau-machines (ICM) basées sur les PEC reposent sur la détection de ces potentiels dans l'électroencéphalogramme (EEG) en réponse à une stimulation planifiée, afin de détecter l'intention de l'utilisateur d'interagir avec une action proposée. Ce type d'interface, imaginé au début des années 1970 par Vidal [1] permet aux personnes paralysées d'interagir avec un ordinateur - l'utilisation de ces interfaces ne nécessitant aucune interaction musculaire (e.g. [2]). Ces interfaces reposent sur la présentation d'une stimulation visuelle attendue mais imprévisible et très distinctive à l'écran (paradigme *oddball*). D'autres paradigmes existent, comme l'imagerie motrice dans laquelle l'utilisateur imagine une action, ou les *steady-state visually evoked potentials* qui consistent en la présentation de stimulations clignotant à différentes fréquences (e.g., [3], [4]). En général, les ICM basées sur le paradigme oddball offrent un bon compromis entre performance, et fatigue de l'utilisateur (notamment visuelle). Bien que nous nous concentrions ici sur la classification des données EEG, obtenues de façon non invasives avec le paradigme oddball, il convient de noter que les techniques de classification sont similaires pour tous les paradigmes.

Les performances des ICM restent limitées pour trois raisons. Tout d'abord, dans le cas de l'EEG non-invasive, les électrodes ne perçoivent qu'un signal de faible amplitude par l'intermédiaire du crâne et restent extrêmement sensibles aux artefacts électromagnétiques et musculaires (induits par exemple par des mouvements ou des lignes de puissance). Ensuite, nous observons un phénomène connu sous le nom d'analphabétisme, qui décrit le fait qu'un pourcentage variant entre 15 et 30% des personnes ne peuvent pas écrire en utilisant les ICM. Le concept d'analphabétisme suggère qu'il s'agit d'un trait physiologique des participants (e.g. [5]), bien que cette conception soit critiquée dans la littérature récente qui pointe plutôt un défaut dans les méthodes de stimulation [6], [7]. Enfin, le taux de transfert d'information de ces interfaces est très faible, avec une vitesse d'écriture d'environ cinq mots par minute [2]. En raison de leur faible performance par rapport aux interfaces mécaniques, les ICM sont donc davantage adaptées au domaine clinique (e.g. [8]), où elles peuvent malgré tout offrir une alternative précieuse aux personnes souffrant de paralysie généralisée. Notez qu'il existe toutefois des expériences d'utilisation des ICM auprès du grand public. Dans ce cas, l'accent est davantage mis sur des caractéristiques comme l'usabilité, la jouabilité ou l'ergonomie, afin de compenser le manque de fiabilité de ces interfaces [9]–[12].

Pour améliorer les performances des ICM, beaucoup d'efforts ont donc été consacrés à la recherche de classifieurs fiables, tels que l'analyse discriminante, la classification par vecteurs de support (SVC pour Support Vector Classification en anglais), les réseau neuronaux [13], les forêts d'arbre décisionnelles [14], ou la géométrie riemannienne [15]. À notre connaissance, les approches s'appuyant sur la géométrie riemannienne, notamment les classifieurs dit MDM (pour Mean Distance to Means en anglais) ont obtenu les meilleures performances lors des compétitions internationales, avec des précisions rapportées proches de 90% en quelques secondes [15]–[17].

En complément de ces études, nous avons étudié l'utilisation de SVC quantiquement améliorés. (QSVC), en utilisant l'implémentation de Havlíček et al. [18], qui est contenue dans la librairie Qiskit (IBM, Armonk, the US) [19]. QSVC est semblable à SVC excepté que l'optimisation quantique est utilisée à deux moments. Dans un premier temps, afin d'estimer le noyau pour toutes les paires de données d'entraînement. Deuxièmement, pour estimer le noyau après l'ajout d'une nouvelle donnée. QSVC a montré des résultats encourageants par rapport à la classification traditionnelle (e.g. [18], [20]). Ceci, associé au développement de services informatiques quantiques disponible sur le cloud (tels que l'expérience quantique IBM d'IBM, Armonk, États-Unis) et avec l'amélioration continue du volume quantique, a fait de la classification quantique une alternative prometteuse, du moins complémentaire, à l'informatique classique.

Dans ce travail, nous avons mesuré la précision équilibrée et le score F1 de QSVC, et les avons comparés aux valeurs obtenues par classifieurs de l'état de l'art, SVC et MDM. Ces deux métriques (précision équilibrée et score F1) sont adaptées pour mesurer la performance d'un jeu de données déséquilibré. Les résultats montrent que QSVC a pu apprendre des données (précision lors de l'entrainement = 83,17 %). Cependant, d'autres recherches sont nécessaires pour améliorer la précision de prédiction du classifieur.

Le présent document est structuré comme suit. La deuxième section décrit les données que nous utilisons pour cette analyse. La troisième section décrit les méthodes que nous avons utilisées pour évaluer la performance de QSVC. La quatrième section contient le résultat de cette expérience. La dernière section contient notre discussion et notre conclusion.

## Données

Nous avons utilisé les données enregistrées au GIPSA-lab (Saint-Martin-d'Hères, France), et disponibles gratuitement sur Zenodo (Genève, Suisse) à partir du lien https://zenodo.org/record/2649069. Le jeu de données contient les enregistrements EEG de 26 participants (7 femmes) âgés en moyenne (std) de 24,4 (2,76) ans, et participant à une expérience visuelle P300 TARGET/NON-TARGET.

Le P300 visuel est un PEC endogène culminant entre 240 et 600 ms après l'apparition d'une stimulation visuelle à l'écran. Contrairement aux composants exogènes à courte latence, qui sont des réponses automatisées et sensorielles à une stimulation, les composants endogènes reflètent le traitement neuronal induit de façon spécifique par la tache réalisée [21]. En particulier, le P300 est suscité par l'apparition d'une stimulation improbable et très distinctive (paradigme oddball).

Les participants jouaient à *Brain Invaders,* une version du célèbre jeu vintage Space *Invaders* (Taito, Tokyo, Japon) pour les ICM. Le jeu est composé de 36 extraterrestres affichés dans une matrice 6x6 (**Figure 1**). La tâche des participants consistait à compter le nombre de flashs d'un extraterrestre TARGET, désigné au début de chaque huit *répétitions.* Dans le paradigme Brain Invaders P300, une répétition est composée de 12 flashs dont deux incluent l'extraterrestre TARGET et 10 ne le font pas (NON-TARGET). Pour chaque participant, il y avait un total de huit extraterrestres TARGET prédéfinis aléatoirement avant expérience. Par conséquent, un total de (resp.) 128 (8x8x2) et 640 (8x8x10) essais TARGET et NON-TARGET ont été enregistrés pour chaque participant au cours de l'expérience.

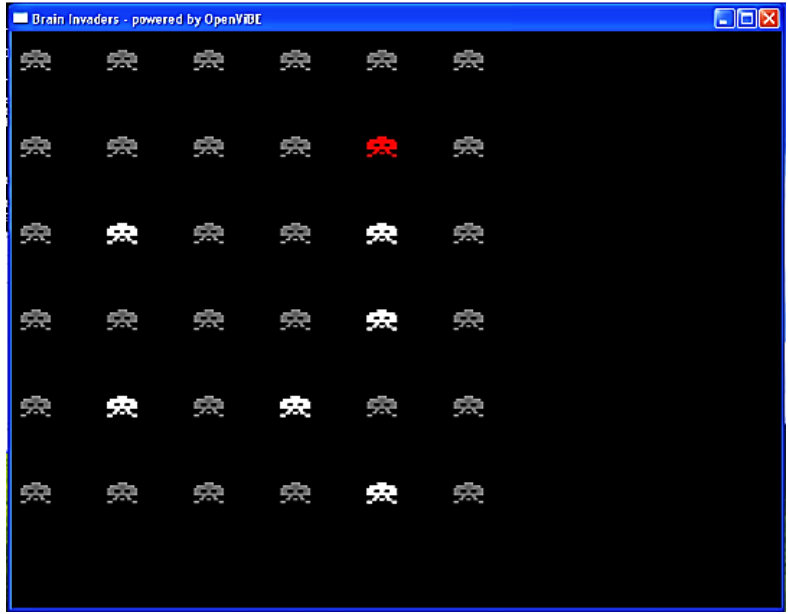

**Figure 1**. Interface de Brain Invaders au moment où un groupe de six symboles NON-TARGET clignote (en blanc). Le symbole rouge est la cible. Les NON-TARGET qui ne clignotent pas sont en gris. (Figure adaptée de Van Veen et al. [22]).

Les signaux EEG ont été acquis au moyen du NeXus-32 (MindMedia, Herten, Allemagne), équipé de 16 électrodes humides, placées selon le système international 10-20 (**Figure 2**). Les signaux ont été enregistrés à une fréquence d'échantillonnage de 128 Hz. Une description complète de l'ensemble de données est disponible dans Van Veen et al. [22].

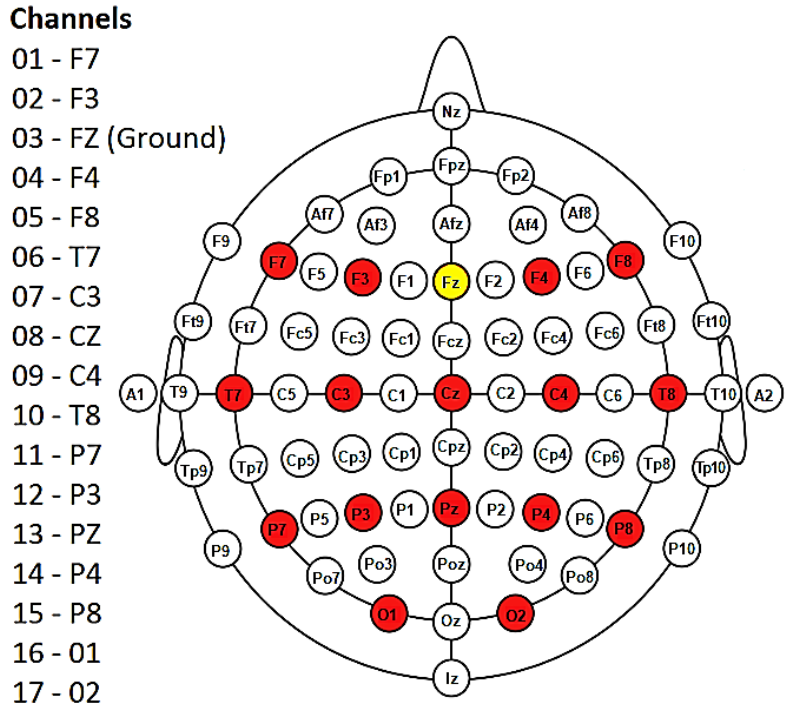

**Figure 2.** En rouge, les 16 électrodes placées selon le système international 10-20. Fz (en jaune) est la terre. Notez que le casque NeXus-32 n'utilise pas d'électrode comme référence, mais qu'une référence moyenne commune et déterminée par le matériel est utilisée. (Figure modifiée à partir de Van Veen et al. [22]).

## Méthode

Les données ont été filtrées entre 1 et 24 Hz à l'aide d'un filtre IR sans phase avec fenêtre de Hamming. Nous avons extrait toutes les époques TARGET (n = 128) et NON-TARGET (n = 640) de 100 à 700 ms après le début d'une stimulation, puis appliqué un filtre spatial xDAWN *(nombre de filtres* = 1) [23]. L'utilisation d'un filtre spatial permet de réduire la dimensionnalité des époques (et donc le temps de calcul), tout en améliorant le rapport signal/bruit. Nous avons transformé ces époques en matrices de corrélation définies positives symétriques (SPD) en utilisant la méthode décrite dans Congedo [16]. Les matrices contenaient 16 éléments (4x4). En considérant la géométrie riemannienne des matrices SPD, nous avons vectorisé ces matrices par projection dans l'espace tangent de la variété riemannienne [24]. Tous les vecteurs contenaient 10 éléments. Les vecteurs résultants ont été donnés en entrée à QSVC. Ensuite, les données ont été intriquées linéairement à l'aide d'un circuit d'évolution Pauli-Z de second ordre (appelé *ZZFeatureMap* dans Qiskit) répétée 2 fois. A chaque classe a été associe un label -1 ou 1, et le kernel a été paramétré avec *enforce_spd=True*, de sorte que les valeurs propres négatives de la matrice du kernel soient remplacées par des zéros. Les autres hyperparamètres étaient le nombre d'itération (500), le facteur de régularisation en norme L2 (0.001) et le nombre d'essais (1024). L'ordinateur quantique a été émulé à l'aide de QasmSimulator [19], et nous avons réalisé une centaine d'exécutions sur un ordinateur quantique IBM avec 16 qubits [25] pour comparer le temps d'exécution.

La performance du classifieur a été évaluée avec une validation croisée a 5 plis et estimée grâce à la métrique dite de précision équilibrée, définie par $\frac{1}{2}(\frac{A}{A+B} + \frac{C}{C+D})$ où A et B (respectivement C et D) représentent le nombre d'époques TARGET (resp. NON-TARGET) correctement et incorrectement classées. De plus, nous avons également calculé le score F1 défini par $\frac{2A}{2A+B+D}$. Les précisions équilibrées et les scores F1 de QSVC ont été comparés à ceux obtenus par les classifieurs SVC et MDM. Pour être objectif dans notre comparaison, la performance de ces deux derniers classifieurs a été mesurée grâce aux mêmes métriques, en utilisant le même jeu de données. Le prétraitement et la validation croisée étaient également similaires. Quand a SVC, il était basé sur un kernel RBF (acronyme anglais pour Radial Basis Function) avec *gamma* fixé à 0,1 – c'est-à-dire l'inverse de la dimension des vecteurs d'entrée (10). Les autres hyperparamètres étaient similaires à ceux de QSVC.

Le traitement du signal a été réalisé grâce à MNE [26], et nous avons utilisé pyRiemann [27] pour le filtrage spatial et la manipulation de matrices de covariance à l'aide de la géométrie Riemannienne. Nous avons utilisé Qiskit [19] pour les opérations quantique. L'évaluation des classifieurs a été réalisée à l'aide de scikit-learn [28].

## Résultats

QSVC a atteint une précision équilibrée moyenne (std) de 83,17 (1,04) %. Comme indiqué dans la **Figure 3**, cette valeur est proche de celle obtenue par SVC et montre que le classifieur a pu apprendre des données.

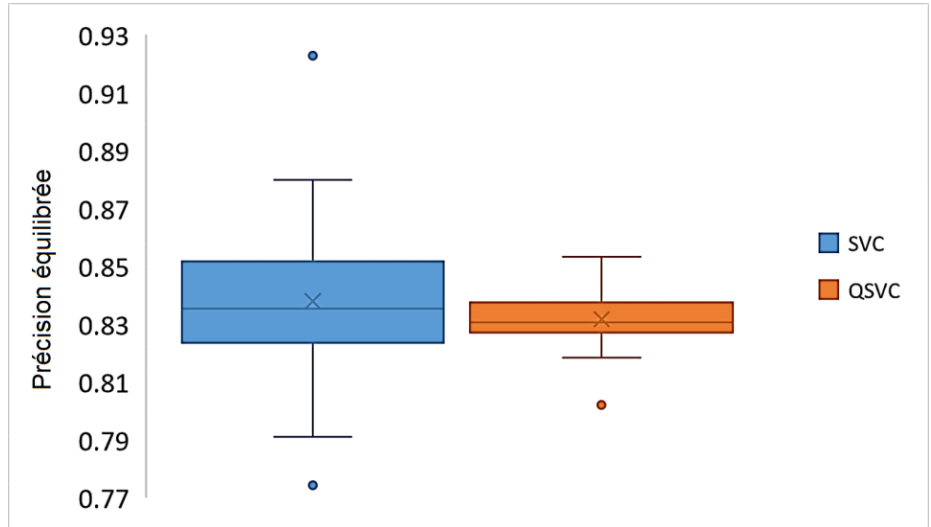

**Figure 3**. Diagramme en boîte des précisions équilibrées obtenues lors de l'entrainement de QSVC et SVC.

La précision équilibrée moyenne (std) et le score F1 obtenus par QSVC lors de la prédiction étaient respectivement de 50,25 (0,83) et 2,84 (1,58). La **Figure 4** comparent ces valeurs avec celles obtenues par MDM et SVC. Une précision équilibrée proche de 50 indique que toutes les époques ont été classées dans la même classe, et un score F1 proche de zéro indique que soit la précision soit le rappel était nul. Pris ensemble, cela suggère probablement que la sensibilité (rappel) était faible.

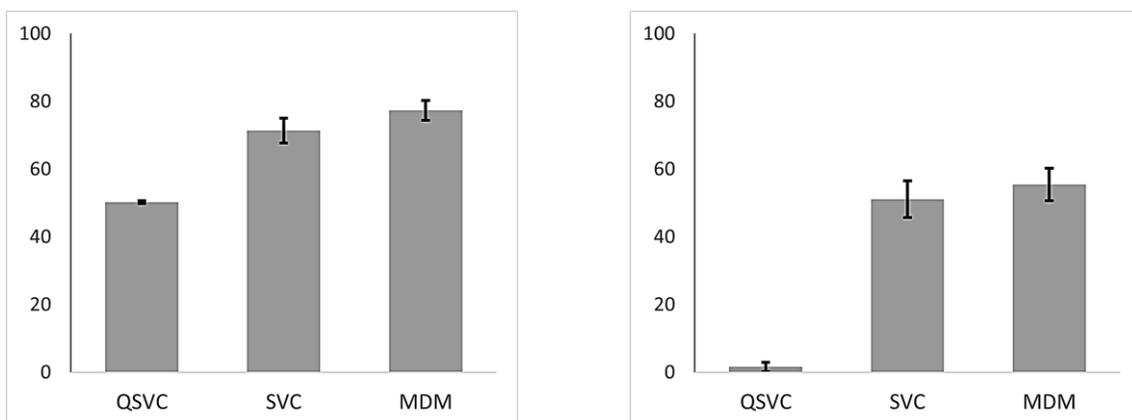

**Figure 4**. Précisions équilibrées (à gauche) et scores F1 (à droite) obtenus lors de la prédiction des classifieurs QSVC, SVC et MDM. Les précisions et les scores sont exprimés en pourcentage.

L'évaluation d'un seul pli a pris environ 8 heures sur notre machine locale, alors qu'il a fallu environ 2 minutes sur le serveur quantique d'IBM. En comparaison, SVC et MDM avaient un temps de classification inferieurs à 2 secondes.

**Discussion and Conclusion**

Les résultats préliminaires présentés dans cet article ont montré que QSVC a atteint une grande précision lors de l'entrainement et a donc pu apprendre des données. Cependant, les performances du classifieur lors de la prédiction étaient faibles par rapport aux résultats présentés dans l'état de l'art. Cela suggère que l'entrainement était trop spécifique et que QSVC nécessite donc un plus grand nombre d'échantillons pour obtenir une meilleure précision de prédiction. Il est également possible que le nombre d'essais ai été trop faible en ce qui concerne la dimension des vecteurs d'entrée (10 dans le cas présent). En fait, une dimension élevée augmente le nombre de mesures possibles, autrement dit le nombre d'erreurs possibles. Une deuxième raison pourrait être que le noyau était trop simple pour obtenir un avantage quantique par rapport à un ordinateur classique. Pour obtenir un avantage sur l'informatique classique, l'encodage des données doit implémenter des circuits quantiques difficiles à émuler sur un ordinateur classique. En pratique, cela pourrait être amélioré en modifiant les paramètres de l'objet *ZZFeatureMap*, comme augmenter la profondeur du circuit (tout en prenant en compte la décohérence du circuit qui augmente avec sa profondeur [18]) ou changer la méthode d'intrication – car nous n'avons utilisé ici que 2 répétitions et intriqué linéairement les données. Cela suggère également que la classification quantique fonctionnera mieux dans les situations où un SVC classique échoue, comme cela peut être le cas dans une expérience avec des personnes présentant un analphabétisme avec les ICM.

Un autre aspect est que le temps d'entrainement et de prédiction étaient significativement plus élevées pour QSVC par rapport aux classifieurs SVC et MDM. Bien que l'informatique quantique soit une technologie prometteuse, cela démontre que les serveurs quantiques ne sont pas encore prêts pour les ICM en utilisation réelle et, en général, pour les applications homme-machines nécessitant une interaction réactive.

Ces résultats doivent être considérés comme une première étape, appelant à d'autres recherches sur la classification quantique appliquée à l'apprentissage supervisé et aux ICM. Par exemple, une piste de recherche comprend l'utilisation d'un classifieur quantique variationnel (Variational Quantum

Classifier en anglais), également présenté dans Havlíček et al. [18]. Une autre orientation de recherche serait d'étudier l'utilisation de l'informatique quantique appliquée au classificateur MDM. Deux études appuient cette suggestion. Tout d'abord, le Manifold Convex Class Model [29], qui est une approche efficace pour transformer les classes de variétés SPD en un modèle convexe, et réaliser la classification en calculant les distances par rapport aux modèles convexes. Deuxièmement, le travail de Apeldoorn et al. [30], qui a démontré que l'informatique quantique appliquée à des problèmes d'optimisation convexe pouvait entraîner une accélération par rapport à l'informatique classique – ce qui à son tour permettrait de diminuer le risque d'erreur en abaissant les seuils de décision.

## Référence